\documentclass[prl,twocolumn,showpacs,preprintnumbers,
		a4paper,superscriptaddress]{revtex4-1}
\usepackage{amsmath}
\usepackage{epsf}
\usepackage{mathtools}
\usepackage{amsfonts}
\usepackage{amssymb}
\usepackage{color}
\usepackage{graphicx}
\usepackage{hyperref}
\def\m@thcombine#1#2{%
  \setbox0=\hbox{$#1$}
  \setbox1=\hbox{$#2$}
  \ifdim\wd0>\wd1
    \setbox0=\hbox to\wd1{\hss\box0\hss}
  \else
    \setbox1=\hbox to\wd0{\hss\box1\hss}
  \fi
  \mathop{\vcenter{
    \offinterlineskip\box0\box1}}}
\def\gesim{\m@thcombine>\sim}
\def\be{\begin{equation}}
\def\ee{\end{equation}}

\begin{document}

\title{Correlations and phase structure of Ising models at complex temperature}
\author{F. \surname{Beichert}}
\affiliation{Scottish Universities Physics Alliance, School of Physics and Astronomy,
University of St Andrews, North Haugh, St Andrews, Fife KY16 9SS, United
Kingdom}
\author{ C. A. \surname{Hooley} }
\affiliation{Scottish Universities Physics Alliance, School of Physics and Astronomy,
University of St Andrews, North Haugh, St Andrews, Fife KY16 9SS, United
Kingdom}
\author{ R. \surname{Moessner} }
\affiliation{Max-Planck-Institut f{\"u}r Physik komplexer Systeme, 01187 Dresden, Germany}
\author{V. \surname{Oganesyan}}
\affiliation{Department of Engineering Science and Physics, College of Staten Island,
CUNY, 2800 Victory Blvd., Staten Island, NY 10314, USA}
\affiliation{The Graduate Center, CUNY, 365 5th Ave., New York, NY 10016}

\begin{abstract}
We investigate the spin-spin correlation functions of Ising magnets at complex values of the temperature, $T$.  For one-dimensional chain and ladder systems, we show the existence of a kind of helimagnetic order in the vicinity of contours where the leading two eigenvalues of the transfer matrix become equal in magnitude.  We analyse the development of long-range order as the two-dimensional limit is approached, and find that there is rich structure in much of the complex-$T$ plane.  In particular, and contrary to the work of Fisher on this problem, the development of long-range order is actually associated with a proliferation of partition function zeros in a certain finite region of that plane containing the real-temperature magnetically ordered phase.  The thermodynamic consequences of this are also discussed.
\end{abstract}

\pacs{05.70.Fh, 64.60.De, 64.60.F-, 75.10.Pq}
\maketitle

{\it Introduction.}  There has recently been a resurgence of interest in the technique of using complex temperature as a probe of the physics of statistical mechanical models \cite{blythe2002,kenna2006,kehrein2012,garrahan2013}, both in and out of equilibrium.  The idea of extending the equilibrium free energy to complex values of the temperature was first introduced by Fisher \cite{fisher1965}, following the related idea of using a complex magnetic field due to Lee and Yang \cite{lee1,lee2}.  However, this has always been viewed as something of a technical trick, notwithstanding occasional statements in the literature such as ``[e]ach region of holomorphy [of the free energy as a function of complex temperature] --- if it comprises at least parts of the real axis --- can be interpreted as a generalized `phase' of the system'' \cite{grossmann1969}.

Given that some of the abovementioned recent work \cite{kehrein2012} explicitly exploits the relationship between real time (relevant to dynamics) and imaginary temperature (as used by Fisher), it may be time to revisit the complex-temperature treatment of statistical mechanical models.  In particular, a more solid interpretation of the physics at complex temperature should be provided.  In what sense is each region of holomorphy a ``generalized `phase' of the system''?  Is the qualification ``if it comprises at least parts of the real axis'' necessary?

In this Letter, we address these questions by examining the correlation functions at complex temperature, and the way in which long-range order in those correlation functions is related to the (more usually studied) distribution of zeros of the partition function.  We do this in detail for a few particular cases:\ the one-dimensional Ising chain, quasi-one-dimensional Ising ladders, and the two-dimensional square-lattice Ising model.  However, we believe that several of our conclusions are more generally applicable; we give arguments to this effect below.

{\it One-dimensional Ising chain.}  Let us begin by briefly analysing the one-dimensional Ising model at complex temperature.  The Hamiltonian is
\be
H = -J \sum_{j=1}^N \sigma_j \sigma_{j+1}, \label{isingchainham}
\ee
where the spin variable on each site $\sigma_j = \pm 1$, and we impose periodic boundary conditions so that $\sigma_{N+1} \equiv \sigma_1$.  The partition function may be very easily obtained using a transfer-matrix method \cite{kramers1941}:
\be
Z = 2^N \cosh^N K + 2^N \sinh^N K, \label{isingchainz}
\ee
where $K \equiv J/k_B T$.  The second term is often dropped because it is sub-dominant in the thermodynamic limit.  However, it is cancellation between these two terms that provides the main structure of the model at complex temperature.  In addition, as we shall see below, the sub-dominant transfer matrix eigenvalues determine the correlation length of the spin-spin correlator.

The conventional approach is now to calculate the values of $K$ for which the partition function (\ref{isingchainz}) is zero; these correspond to singular points of the free energy, and thus are related to phase transitions.  Clearly (\ref{isingchainz}) is zero when
\be
\left( \tanh K \right)^N = -1,
\ee
or, taking the $N$th root,
\be
\tanh K = e^{i\theta_n},
\ee
where the angles $\theta_n$ specify $N$ equally-spaced points on the unit circle:
\be
\theta_n = \frac{2\pi}{N} \left( n + \frac{1}{2} \right), \quad n = 0,1,2,\ldots,N-1.
\ee
Hence the zeros of this partition function all lie on the unit circle in the complex $\tanh K$ plane.  Since this variable is clearly the natural one when studying partition function zeros in Ising problems, we shall henceforth define $z \equiv \tanh K$.  We shall sometimes, for brevity, refer to this variable as the temperature, even though really it only encodes it.

Non-analyticities in the free energy, i.e.\ phase transitions, can occur in general when we move from a region of the $z$-plane in which one transfer-matrix eigenvalue is dominant to a region in which a different one is.  Hence, at a phase transition, we expect the two leading eigenvalues of the transfer matrix to become degenerate.  This criterion, however, is deeply related to the existence of zeros of the partition function.  Indeed, there is a theorem due to Beraha, Kahane, and Weiss \cite{beraha1975}, which essentially states that (except for certain isolated points) functions of the form
\be
\sum_j \alpha_j \lambda_j^N
\ee
can have zeros only at points where $\vert \lambda_1 \vert = \vert \lambda_2 \vert$, where the eigenvalues $\lambda_j$ are numbered in order of their magnitudes with the largest first.  Hence any phase transition must coincide with the crossing of a contour of partition function zeros.

To characterise such transitions, we consider the latent heat.  This can be evaluated as the difference in the internal energy per particle between a point just `inside' the unit circle and a point just `outside' it.  Continuing the usual thermodynamic formulas to complex values of the temperature, it is easy to show that
\be
\frac{\Delta E}{N} \equiv \frac{E_{\rm out}}{N} - \frac{E_{\rm in}}{N} = \lambda_{\rm in}^{-1} \frac{\partial \lambda_{\rm in}}{\partial \beta} - \lambda_{\rm out}^{-1} \frac{\partial \lambda_{\rm out}}{\partial \beta}, \label{latheat}
\ee
where $\lambda_{\rm in}$ is the transfer matrix eigenvalue that is dominant inside the unit circle ($2\cosh K$ in this case), $\lambda_{\rm out}$ is the one that is dominant outside ($2\sinh K$), and $\beta \equiv 1/k_B T$.  Evaluating this latent heat at a temperature $z=e^{i\theta}$ on the unit circle, we obtain
\be
\frac{\Delta E}{N} = 2i J \sin\theta.
\ee
We see that, for real temperatures ($\theta = 0$ or $\theta=\pi$) there is no latent heat, and the transition is continuous; whereas for any other temperature the transition is first-order, with a purely imaginary latent heat.

As well as determining the locations of the partition function zeros and of the phase transitions, eigenvalue degeneracies also determine the temperatures at which one sees long-range order in the correlation functions.  This is because, in general, correlation functions are expressed in terms of transfer-matrix eigenvalues as
\be
\langle \sigma_0 \sigma_j \rangle = \gamma_2 \left( \frac{\lambda_2}{\lambda_1} \right)^j + \gamma_3 \left( \frac{\lambda_3}{\lambda_1} \right)^j + \ldots . \label{correl}
\ee
For this to be non-zero as $j \to \infty$, it must be the case that $\vert \lambda_2 \vert = \vert \lambda_1 \vert$, which is also the criterion for the possibility of a partition function zero (as per Beraha-Kahane-Weiss) and the existence of a non-analyticity in the free energy (see above).
\begin{figure}
\begin{center}
\includegraphics[width=8.5cm]{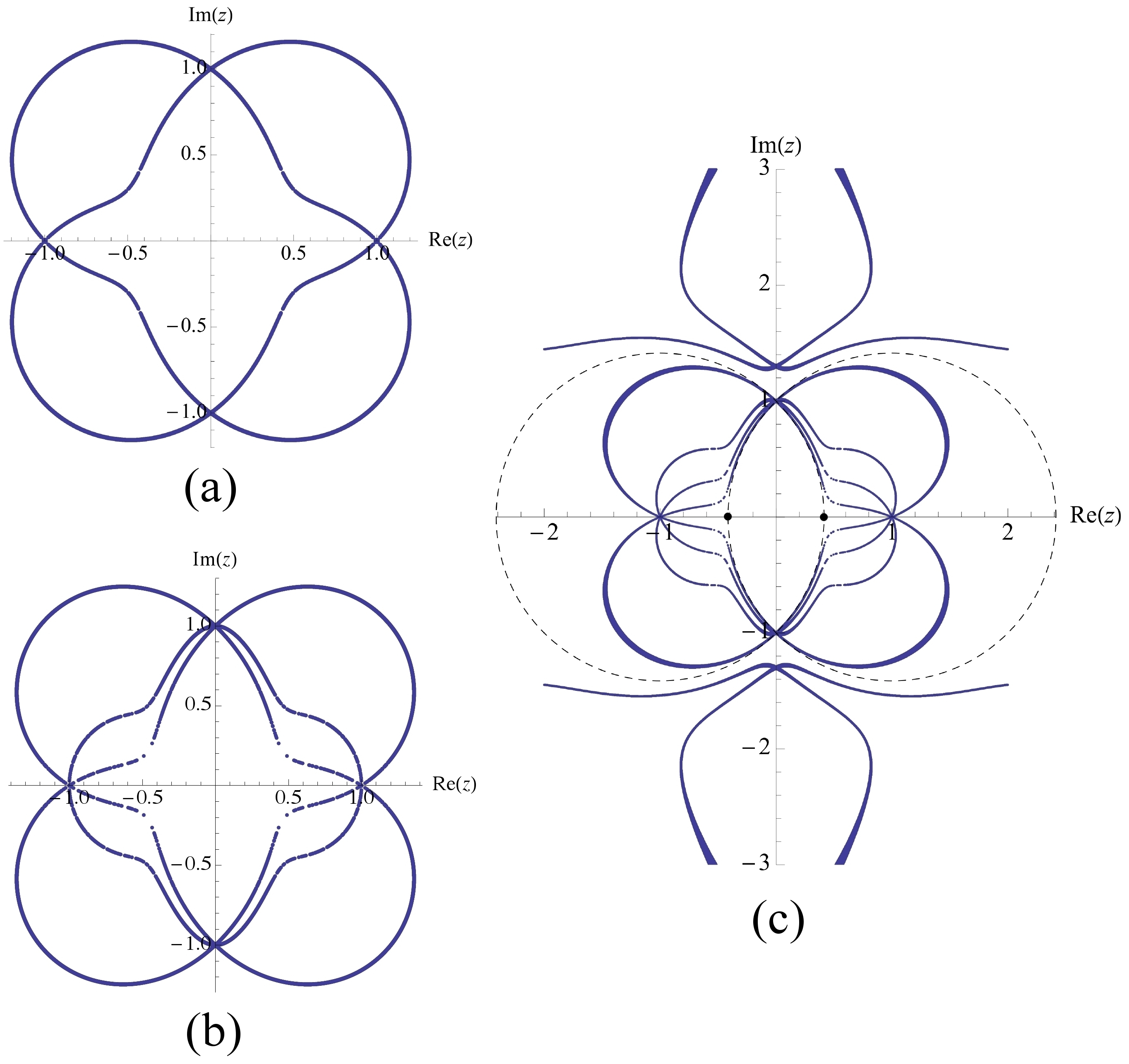}
\end{center}
\caption{Contours in the complex $\tanh (J/k_B T)$ plane along which the two leading transfer-matrix eigenvalues are degenerate for Ising ladders of various widths: (a) two legs; (b) three legs; (c) four legs.  The dashed lines in panel (c) show, for comparison, the `Fisher circles' \cite{fisher1965} on which the partition function zeros of the isotropic two-dimensional Ising model lie.  The solid circles on the real-temperature axis correspond to the ordering transitions of the isotropic two-dimensional Ising model:\ antiferromagnetic at $z=1-\sqrt{2}$; ferromagnetic at $z=\sqrt{2}-1$.}
\label{ladderdeg}
\end{figure}

Notice, however, that (\ref{correl}) is also sensitive to the relative phase of the two largest eigenvalues.  In the Ising chain, this gives rise to {\it helimagnetic\/} correlations when the temperature is complex.  To see this, consider a temperature $z=R e^{i\theta}$ which lies near the unit circle.  Explicit calculation gives
\be
C_j \equiv \langle \sigma_0 \sigma_j \rangle = \frac{R^j e^{ij\theta} + R^{N-j} e^{i(N-j)\theta}}{1+R^N e^{iN\theta}}.
\ee
For temperatures inside the unit circle, $R$ is slightly less than one, so that for $N \gg j$ we obtain $C_j \sim R^j e^{ij\theta}$.  Outside the unit circle, $R$ is slightly greater than one, so that $C_j \sim R^{-j} e^{-ij\theta}$.  In both cases the correlations decay exponentially with distance, though the correlation length diverges as the temperature approaches the unit circle.  But the correlations also have a spiral nature, and the handedness of this spiral switches as the circle is crossed.  On the unit circle itself, the two spirals combine to form a density wave,
\be
C_j \sim \cos \left( \theta j \right). \label{densitywave}
\ee
Notice that this density wave evolves smoothly from uniform long-range order at the ferromagnetic point ($\theta=0$) to N{\'e}el long-range order at the antiferromagnetic point ($\theta=\pi$) as the unit circle is traversed.  Furthermore, the correlation functions are less sensitive to the boundary conditions of the problem than the contours of partition function zeros are, since the correlation functions sample the sub-dominant eigenvalue of the transfer matrix even if that eigenvalue is accidentally absent from the partition function.

One of the features of the treatment of Ising models at complex temperature is that the $Z_2$ spins $\sigma_j$ are effectively promoted to $U(1)$ (i.e.\ XY-type) spins.  This is not through a change in the nature of the partition sum, however, but through the fact that the statistical weights used in calculating the correlation functions are now complex.  Hence it is strictly expectation values, not the spins themselves, that have their symmetry group enlarged.  However, much of the intuition of spiral magnetism of XY-spins at real temperature carries across to this complex-temperature Ising case.

{\it Ising ladders.} It is natural to wonder which, if any, of these features survive into higher dimensions.  To address this question, we now consider repeating the above analysis for the case of Ising ladders.  Consider a ladder with $N_L$ legs, and $N \gg N_L$ sites on each leg.  We take periodic boundary conditions in the `long' direction, but open boundary conditions in the `short' direction.  For any number of legs greater than two, the problem already becomes analytically intractable, so most of our results in this section are numerical.  We shall first present these, and then give an analytical derivation of some of the observed properties near the real-temperature line.

Fig.~\ref{ladderdeg} shows the contours along which $\vert \lambda_1 \vert = \vert \lambda_2 \vert$ for Ising ladders with two, three, and four legs.  These are determined numerically, though of course for the $N_L=2$ case analytical calculation is also possible and agrees with the numerical results.  In Fig.~\ref{ladderdeg}(c), as well as showing an expanded region of the complex-temperature plane, we have added the two circles along which Fisher \cite{fisher1965} determined the partition-function zeros of the isotropic two-dimensional Ising model to lie.

Note the regular structure that appears around the points $z=1$ and $z=-1$, which correspond respectively to the ferromagnetically and antiferromagnetically ordered $T=0$ states of the ladder.  It seems from the figure that, for a ladder with $N_L$ legs, precisely $2N_L$ degeneracy contours meet at each of these points, and that those contours are equally spaced in angle.  This can in fact be proved as follows.  Consider making a low-temperature expansion of the partition function around the point $z=1$.  The two dominant configurations of the system are the two ferromagnetic ground states, which are present at $T=0$.  As further terms in the expansion are included, these will dress the two ground states; but they will dress both of them in the same way, so the degeneracy between their contributions to the partition function will be maintained.  The first excitation that breaks that degeneracy is a domain wall extending across the ladder in the short direction.  Its effect may be described by the reduced transfer matrix
\be
{\bf T} = \left( \begin{array}{cc}
1 & w^{N_L} \\
w^{N_L} & 1 \end{array} \right).
\ee
The entries on the diagonal are the contributions of the two ground states to the (renormalised) partition function; each off-diagonal entry is the weight corresponding to a domain wall, where $w = e^{-2J/k_B T}$ is the weight corresponding to an unsatisfied bond.  The eigenvalues of this matrix are
\be
\lambda_{\pm} = 1 \pm w^{N_L},
\ee
and these are the dominant two eigenvalues of the full transfer matrix.  They become degenerate when $\vert \lambda_{+} \vert = \vert \lambda_{-} \vert$; it is easily shown that this condition is satisfied when
\be
z-1 = R \exp \left[ \frac{i\pi}{N_L} \left( n + \frac{1}{2} \right) \right], \quad n=1,2,\ldots,2N_L, \label{crossings}
\ee
i.e.\ that $2N_L$ equally spaced degeneracy contours converge at the point $z=1$.  The analogous result for the $z=-1$ point may be established by very similar arguments.  Note that we must have $R \ll 1$ for the low-temperature expansion to remain valid.

\begin{figure}
\begin{center}
\includegraphics[width=8.5cm]{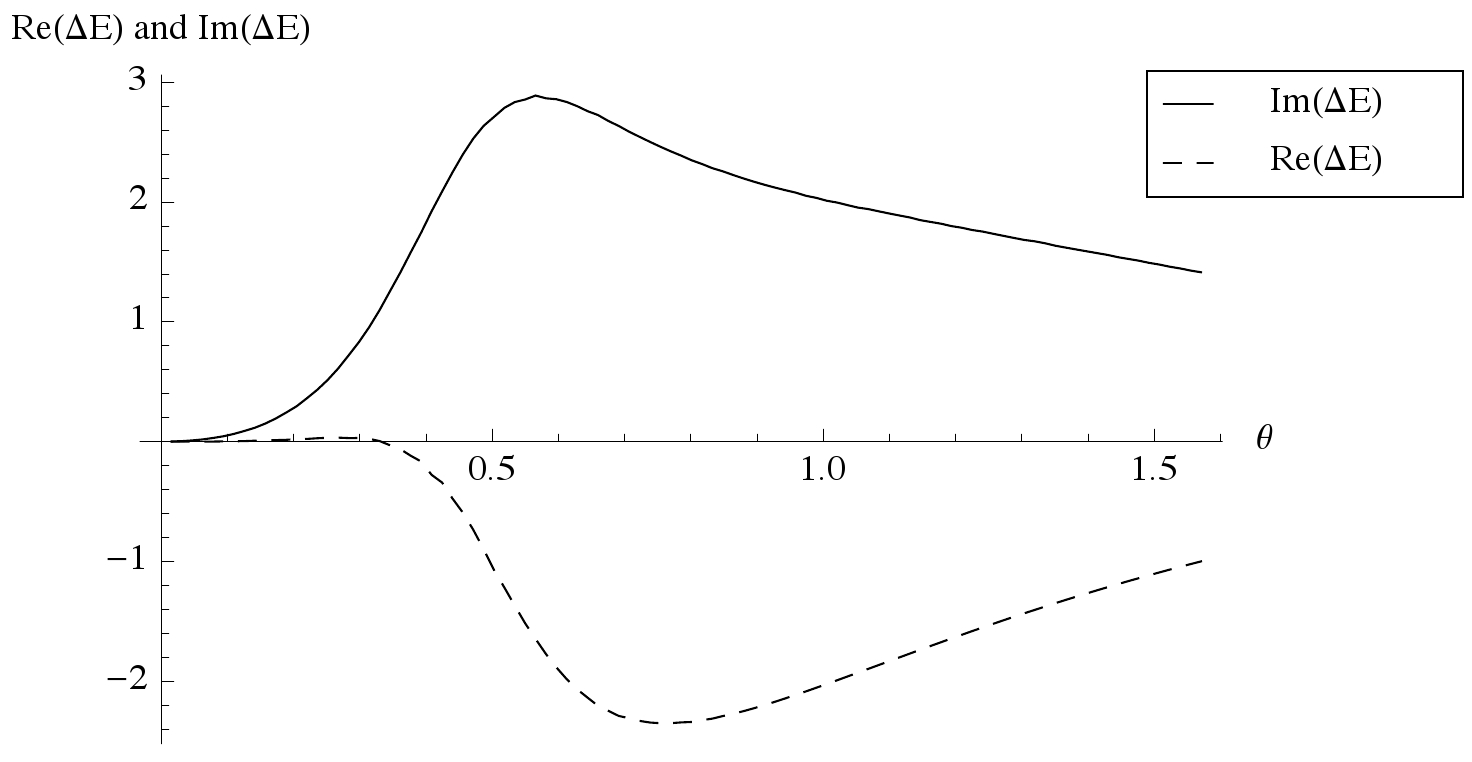}
\end{center}
\caption{The real and imaginary parts of the latent heat per spin when the phase transition corresponding to the inner contour of Fig.~\ref{ladderdeg}(a) is crossed via a radial path in the complex temperature plane at an angle $\theta$ to the real line.  Notice that the latent heat is purely imaginary (as in the Ising chain case) until $\theta$ reaches a critical angle corresponding to the kink in the contour in Fig.~\ref{ladderdeg}(a); after this a finite real part develops.  We conjecture that this is a signature of a crossover from one- to two-dimensional behaviour.}
\label{ladderlatent}
\end{figure}

We can also use this low-temperature expansion to explore the latent heat associated with crossing each of these contours.  Using (\ref{latheat}), we obtain
\be
\frac{\Delta E}{NN_L} = iJ \left( 2 - Re^{i\theta} \right) R^{N_L} (-1)^n \frac{2}{1+R^{2N_L}}, \label{latentladdereq}
\ee
where $\theta$ is one of the angles given in (\ref{crossings}).  Notice that the latent heat vanishes as the $d=2$ Ising limit ($N_L \to \infty$) is taken.  These observations continue to hold true even quite far from the real-temperature line:\ Fig.~\ref{ladderlatent} shows the real and imaginary parts of the latent heat as the inner contour of Fig.~\ref{ladderdeg}(a) is crossed in an outward radial direction at angle $\theta$ to the real line.

\begin{figure}
\begin{center}
\includegraphics[width=8.5cm]{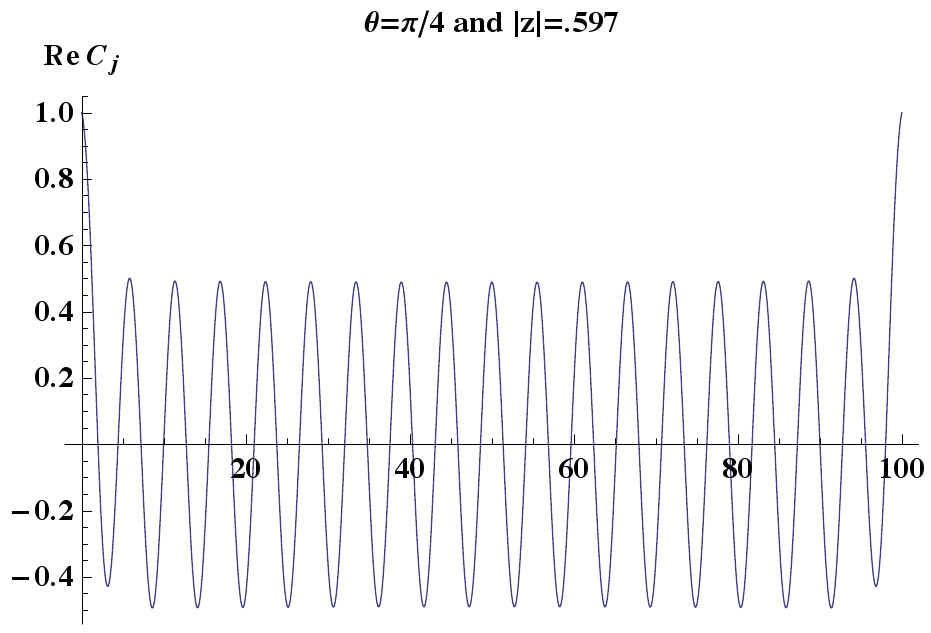}
\end{center}
\caption{The spin-spin correlation function $\langle \sigma_0 \sigma_j \rangle$ for two spins on the same leg of a two-leg ladder with $N=100$ spins on each leg.}
\label{laddercorrel}
\end{figure}

We may also ask about the correlation functions in the ladder case.  Fig.~\ref{laddercorrel} shows the correlation function $C_j$ for two spins on the same leg of the ladder with $N_L=2$ and $N=100$, separated in the leg direction by a distance $j$.  We see that the density-wave behaviour of (\ref{densitywave}) persists, albeit with reduced amplitude, suggesting that our interpretation in terms of effective helimagnetism continues to apply in the ladder case.

{\it Approaching the two-dimensional Ising model.}  The above results give intriguing insights into the approach to the two-dimensional Ising model as $N_L \to \infty$.  Following the pattern of the preceding two sections, let us ask four questions:  First, what happens to the eigenvalue-degeneracy contours in the $N_L \to \infty$ limit?  Second, what happens to the partition function zeros?  Third, what happens to the latent heat?  And fourth, what happens to the correlation functions?

We can confidently answer the first question near the zero-temperature point, since it follows from (\ref{crossings}) that the region around $z=1$ fills in entirely with such contours.  Since the only significant lines in the complex temperature plane of the two-dimensional Ising model in the thermodynamic limit are the Fisher circles, it seems reasonable to conjecture that the entire region inside one Fisher circle but outside the other fills in likewise.  This would imply that $\vert \lambda_1 \vert = \vert \lambda_2 \vert$ in this entire region in the $N_L \to \infty$ limit.

The answer to the second question involves a subtlety in the Beraha-Kahane-Weiss theorem:\ while degeneracy of eigenvalues is necessary for the existence of partition function zeros, it is not sufficient.  In particular, if the condition $\vert \lambda_1 \vert = \vert \lambda_2 \vert$ is obeyed over a finite region in the $z$-plane, it does not lead to zeros of the partition function.  But we do learn that the $N_L \to \infty$ limit is somewhat singular in this regard.

One might still ask, however, why Fisher's method \cite{fisher1966} based on Kasteleyn's solution of the dimer-covering problem \cite{kasteleyn1961} does not see the extra partition function zeros in the case of a two-dimensional lattice of finite extent.  There is nothing in the Fisher-Kasteleyn method that relies on the thermodynamic limit, so in principle all transfer matrix eigenvalues are there.  We believe that the resolution of this conundrum is related to the summation over different topological sectors that is required in the treatment of the two-dimensional Ising model.  Fisher \cite{fisher1965} in fact uses the partition function from only one of these sectors to determine the locations of the circles shown in Fig.~\ref{ladderdeg}(c).  We conjecture that cancellations between the partition functions of the different sectors give the extra zeros that we find in the finite-size case.

Regarding the third question, the formula (\ref{latentladdereq}) strongly suggests that, as the eigenvalue-degeneracy contours proliferate, the latent heat associated with each one is reduced in magnitude (while retaining, in general, both real and imaginary parts).  Indeed, we can see from (\ref{latentladdereq}) that even the sum of $\Delta E/NN_L$ over all the transitions tends to zero as $N_L \to \infty$, which implies that in the vicinity of the $z=1$ point the thermodynamic behaviour becomes entirely regular.  The only thermodynamic discontinuity left is on the Fisher circles themselves, where the transition is presumably first-order.

Finally, what of the correlation functions?  There is nothing in the results obtained above to suggest that these revert to being purely ferromagnetic as the $N_L \to \infty$ limit is taken; and on the other hand there is plenty of evidence \cite{beichert2013} to suggest that they retain a helimagnetic character.  We conjecture that this is a general property of the correlation functions of Ising magnets at complex temperature:\ helimagnetism, with a pitch determined by the position of the temperature on the dense limit of the `co-ordinate grid' of contours seen in Fig.~\ref{ladderdeg}.

This conjecture provides a natural framework in which to understand the instantaneous fragmentation of the region into multiple separated phases as the lattice is made finite in one direction:\ it is a commensuration effect.  Similarly, if one had a spiral magnet with a continuously evolving ${\bf q}$-vector, finite-size effects would force the ${\bf q}$-vector instead to make a sequence of transitions between the values permitted by the boundary conditions.

Direct numerical or analytical evidence for these helimagnetic correlations at complex temperature would be extremely useful.  We also hope that these observations may help in understanding the link between complex temperature and real-time dynamical problems.

{\it Conclusion.}  At the beginning of this Letter, we asked whether regions of holomorphy in the complex plane can properly be interpreted as generalized `phases'.  The answer for one- and quasi-one-dimensional systems is clearly no.  Although these regions exist, long-range order is possible only on the degeneracy contours, which for unfrustrated systems are generally isolated lines.  (For the case of frustrated models, see \cite{beichert2013}.)  In two-dimensional models we have two different types of holomorphic region in the complex-temperature plane:\ type I regions were holomorphic even in the quasi-one-dimensional version of the model, while type II regions became holomorphic via the proliferation and coalescence of degeneracy contours.  In type II regions, we see helimagnetic order with a ${\bf q}$-vector which evolves continuously as the region is traversed.  The intersection of a type II region with the real axis corresponds to long-range order in the real-temperature model.  Further work on this interesting classification of complex-temperature phases is desirable.

This research was supported in part by the US National Science Foundation under DMR-0955714 (VO).
CAH also gratefully acknowledges financial support from the EPSRC (UK) via grants EP/I031014/1 and EP/H049584/1.
\bibliography{paper}

\begin{thebibliography}{13}
\expandafter\ifx\csname natexlab\endcsname\relax\def\natexlab#1{#1}\fi
\expandafter\ifx\csname bibnamefont\endcsname\relax
  \def\bibnamefont#1{#1}\fi
\expandafter\ifx\csname bibfnamefont\endcsname\relax
  \def\bibfnamefont#1{#1}\fi
\expandafter\ifx\csname citenamefont\endcsname\relax
  \def\citenamefont#1{#1}\fi
\expandafter\ifx\csname url\endcsname\relax
  \def\url#1{\texttt{#1}}\fi
\expandafter\ifx\csname urlprefix\endcsname\relax\def\urlprefix{URL }\fi
\providecommand{\bibinfo}[2]{#2}
\providecommand{\eprint}[2][]{\url{#2}}

\bibitem[{\citenamefont{Blythe and Evans}(2002)}]{blythe2002}
\bibinfo{author}{\bibfnamefont{R.~A.} \bibnamefont{Blythe}} \bibnamefont{and}
  \bibinfo{author}{\bibfnamefont{M.~R.} \bibnamefont{Evans}},
  \bibinfo{journal}{Phys. Rev. Lett.} \textbf{\bibinfo{volume}{89}},
  \bibinfo{pages}{080601} (\bibinfo{year}{2002}).

\bibitem[{\citenamefont{Kenna et~al.}(2006)\citenamefont{Kenna, Johnston, and
  Janke}}]{kenna2006}
\bibinfo{author}{\bibfnamefont{R.}~\bibnamefont{Kenna}},
  \bibinfo{author}{\bibfnamefont{D.~A.} \bibnamefont{Johnston}},
  \bibnamefont{and} \bibinfo{author}{\bibfnamefont{W.}~\bibnamefont{Janke}},
  \bibinfo{journal}{Phys. Rev. Lett.} \textbf{\bibinfo{volume}{96}},
  \bibinfo{pages}{115701} (\bibinfo{year}{2006}).

\bibitem[{\citenamefont{Heyl et~al.}(2012)\citenamefont{Heyl, Polkovnikov, and
  Kehrein}}]{kehrein2012}
\bibinfo{author}{\bibfnamefont{M.}~\bibnamefont{Heyl}},
  \bibinfo{author}{\bibfnamefont{A.}~\bibnamefont{Polkovnikov}},
  \bibnamefont{and} \bibinfo{author}{\bibfnamefont{S.}~\bibnamefont{Kehrein}},
  \bibinfo{journal}{arXiv:1206.2505}  (\bibinfo{year}{2012}).

\bibitem[{\citenamefont{Flindt and Garrahan}(2013)}]{garrahan2013}
\bibinfo{author}{\bibfnamefont{C.}~\bibnamefont{Flindt}} \bibnamefont{and}
  \bibinfo{author}{\bibfnamefont{J.~P.} \bibnamefont{Garrahan}},
  \bibinfo{journal}{Phys. Rev. Lett.} \textbf{\bibinfo{volume}{110}},
  \bibinfo{pages}{050601} (\bibinfo{year}{2013}).

\bibitem[{\citenamefont{Fisher}(1965)}]{fisher1965}
\bibinfo{author}{\bibfnamefont{M.~E.} \bibnamefont{Fisher}},
  \emph{\bibinfo{title}{{in: Lectures in Theoretical Physics}}}, vol.
  \bibinfo{volume}{VII C} (\bibinfo{publisher}{Gordon and Breach},
  \bibinfo{year}{1965}).

\bibitem[{\citenamefont{Yang and Lee}(1952)}]{lee1}
\bibinfo{author}{\bibfnamefont{C.~N.} \bibnamefont{Yang}} \bibnamefont{and}
  \bibinfo{author}{\bibfnamefont{T.~D.} \bibnamefont{Lee}},
  \bibinfo{journal}{Phys. Rev.} \textbf{\bibinfo{volume}{87}},
  \bibinfo{pages}{404} (\bibinfo{year}{1952}).

\bibitem[{\citenamefont{Lee and Yang}(1952)}]{lee2}
\bibinfo{author}{\bibfnamefont{T.~D.} \bibnamefont{Lee}} \bibnamefont{and}
  \bibinfo{author}{\bibfnamefont{C.~N.} \bibnamefont{Yang}},
  \bibinfo{journal}{Phys. Rev.} \textbf{\bibinfo{volume}{87}},
  \bibinfo{pages}{410} (\bibinfo{year}{1952}).

\bibitem[{\citenamefont{Grossmann}(1969)}]{grossmann1969}
\bibinfo{author}{\bibfnamefont{S.}~\bibnamefont{Grossmann}},
  \bibinfo{journal}{Festk{\"o}rperprobleme} \textbf{\bibinfo{volume}{9}},
  \bibinfo{pages}{207} (\bibinfo{year}{1969}).

\bibitem[{\citenamefont{Kramers and Wannier}(1941)}]{kramers1941}
\bibinfo{author}{\bibfnamefont{H.~A.} \bibnamefont{Kramers}} \bibnamefont{and}
  \bibinfo{author}{\bibfnamefont{G.~H.} \bibnamefont{Wannier}},
  \bibinfo{journal}{Phys. Rev.} \textbf{\bibinfo{volume}{60}},
  \bibinfo{pages}{252} (\bibinfo{year}{1941}).

\bibitem[{\citenamefont{Beraha et~al.}(1975)\citenamefont{Beraha, Kahane, and
  Weiss}}]{beraha1975}
\bibinfo{author}{\bibfnamefont{S.}~\bibnamefont{Beraha}},
  \bibinfo{author}{\bibfnamefont{J.}~\bibnamefont{Kahane}}, \bibnamefont{and}
  \bibinfo{author}{\bibfnamefont{N.}~\bibnamefont{Weiss}},
  \bibinfo{journal}{Proc. Nat. Acad. Sci. USA} \textbf{\bibinfo{volume}{72}},
  \bibinfo{pages}{4209} (\bibinfo{year}{1975}).

\bibitem[{\citenamefont{Fisher}(1966)}]{fisher1966}
\bibinfo{author}{\bibfnamefont{M.~E.} \bibnamefont{Fisher}},
  \bibinfo{journal}{J. Math. Phys.} \textbf{\bibinfo{volume}{7}},
  \bibinfo{pages}{1776} (\bibinfo{year}{1966}).

\bibitem[{\citenamefont{Kasteleyn}(1961)}]{kasteleyn1961}
\bibinfo{author}{\bibfnamefont{P.~W.} \bibnamefont{Kasteleyn}},
  \bibinfo{journal}{Physica} \textbf{\bibinfo{volume}{27}},
  \bibinfo{pages}{1209} (\bibinfo{year}{1961}).

\bibitem[{\citenamefont{Beichert et~al.}(2013)\citenamefont{Beichert, Hooley,
  Moessner, and Oganesyan}}]{beichert2013}
\bibinfo{author}{\bibfnamefont{F.}~\bibnamefont{Beichert}},
  \bibinfo{author}{\bibfnamefont{C.~A.} \bibnamefont{Hooley}},
  \bibinfo{author}{\bibfnamefont{R.}~\bibnamefont{Moessner}}, \bibnamefont{and}
  \bibinfo{author}{\bibfnamefont{V.}~\bibnamefont{Oganesyan}},
  \bibinfo{journal}{in preparation}  (\bibinfo{year}{2013}).

\end{thebibliography}

\end{document}